\documentclass[12pt]{article}
\usepackage{graphicx}
\usepackage{epsfig}
\usepackage{epstopdf}
\DeclareGraphicsExtensions{.pdf,.eps,.png,.jpg,.mps}

\def\lsim{\mathrel{\rlap{\lower3pt\hbox{\hskip0pt$\sim$}}
     \raise1pt\hbox{$<$}}}         
\def\gsim{\mathrel{\rlap{\lower4pt\hbox{\hskip1pt$\sim$}}
     \raise1pt\hbox{$>$}}}         

\usepackage{amsmath}
\usepackage{amsfonts}

\begin{document}
\begin{titlepage}

\centerline{\Large \bf Russian-Doll Risk Models}
\medskip

\centerline{Zura Kakushadze$^\S$$^\dag$$^\sharp$\footnote{\, \tt Email: zura@quantigic.com}}

\bigskip

\centerline{\em $^\S$ Quantigic$^\circledR$ Solutions LLC}
\centerline{\em 1127 High Ridge Road \#135, Stamford, CT 06905\,\,\footnote{\, DISCLAIMER: This address is used by the corresponding author for no
purpose other than to indicate his professional affiliation as is customary in
publications. In particular, the contents of this paper
are not intended as an investment, legal, tax or any other such advice,
and in no way represent views of Quantigic$^\circledR$ Solutions LLC,
the website \underline{www.quantigic.com} or any of their other affiliates.
}}
\centerline{\em $^\dag$ Free University of Tbilisi, Business School \& School of Physics}
\centerline{\em 240, David Agmashenebeli Alley, Tbilisi, 0159, Georgia}
\centerline{\em $^\sharp$ Institute for Advanced Study}
\centerline{\em Hong Kong University of Science and Technology, Hong Kong}
\medskip
\centerline{(December 14, 2014; revised: March 15, 2015)}

\bigskip
\medskip

\begin{abstract}
{}We give a simple explicit algorithm for building multi-factor risk models. It dramatically reduces the number of or altogether eliminates the risk factors for which the factor covariance matrix needs to be computed. This is achieved via a nested ``Russian-doll" embedding: the factor covariance matrix itself is modeled via a factor model, whose factor covariance matrix in turn is modeled via a factor model, and so on. We discuss in detail how to implement this algorithm in the case of (binary) industry classification based risk factors ({\em e.g.}, ``sector $\rightarrow$ industry $\rightarrow$ sub-industry"), and also in the presence of (non-binary) style factors. Our algorithm is particularly useful when long historical lookbacks are unavailable or undesirable, {\em e.g.}, in short-horizon quant trading.
\end{abstract}
\medskip
\end{titlepage}

\newpage

\section{Introduction and Summary}

{}The basic idea behind multi-factor risk models for equities is that, in lieu of computing a sample covariance matrix (SCM) for a large number of stocks, it allows to compute a factor covariance matrix (FCM) for many fewer risk factors. Why is this useful? When the number of stocks $N$ in the trading universe is large, it is either impracticable or impossible to reliably compute SCM. In practical applications, the number of historical observations in the time series of returns based on which SCM is computed is too low, so SCM is either singular or its off-diagonal elements are not out-of-sample stable. This is further exacerbated by the fact that, for applications involving shorter-horizon quantitative trading strategies, the desirable lookback for the risk model should not be too long. Indeed, for extremely ephemeral short-holding alphas whose lifespan can be as short as a few months, it is not too relevant what SCM matrix looked like, say, 5 years ago -- and 5 years contain only about 1260 daily observations, whereas a typical quantitative trading universe can consist of as many as 2,000-2,500 liquid enough to trade stocks.

{}Multi-factor risk models alleviate this issue by modeling off-diagonal elements of SCM as owing to a much smaller number $K \ll N$ of risk factors, and in the zeroth approximation one can compute sample $K\times K$ FCM. In most common incarnations, these risk factors consist of a combination of style factors and industry (classification based) factors. The number of style factors, which are based on some estimated (or measured) properties of stocks (such as size, momentum, volatility, liquidity, value, growth, {\em etc}.), typically is limited, of order 10 or fewer. However, the number of industry factors can be much larger, from several dozen for less granular incarnations, to a few hundred for more granular risk models, where the number of industry factors can depend on the trading universe. In these cases, if the desirable lookback is short (say, 1-2 years or less), then reliably computing sample $K\times K$ FCM itself becomes impracticable or impossible. What can we do in such cases?\footnote{\, This is especially pertinent to shorter holding horizon {\em quantitative trading} strategies, where, in practice, the relevant lookbacks are short and the number of (industry) risk factors is large (including, in order to achieve higher Sharpe ratios). In contrast, for longer holding horizon strategies (with low Sharpe ratios) the literature typically focuses on a few risk factors, see, {\em e.g.},
(Ang {\em et al}, 2006),
(Anson, 2013/14),
(Asness, 1995),
(Asness and Stevens, 1995),
(Asness {\em et al}, 2001),
(Banz, 1981),
(Basu, 1977),
(Burmeister and Wall, 1986),
(Chen, {\em et al}, 1986, 1990),
(Fama and French, 1992, 1993, 2015),
(Haugen, 1995),
(Jegadeesh and Titman, 1993),
(Lakonishok {\em et al}, 1994),
(Liew and Vassalou, 2000),
(Pastor and Stambaugh, 2003),
(Scholes and Williams, 1977).\label{foot3}}

{}The idea we advance in this paper is quite simple. If sample FCM cannot be computed, we can model FCM itself via a factor model with a much smaller number $F \ll K$ of yet new risk factors. If sample FCM can be reliably computed for these new factors -- good; if not, then we model FCM for the new factors via yet another factor model with a much smaller number $L\ll F$ of yet new risk factors. And so on -- until either we can reliably compute FCM for the resulting (small enough number of) risk factors, or we altogether eliminate the need for computing non-diagonal FCM by reducing the number of risk factors to 1 (in this case we have $1\times 1$ FCM, which is a variance and can be reliably computed) or even 0. We refer to this construction as a (nested)\footnote{\, In a different context, Chicheportiche and Bouchaud (2014) use a nested factor approach to model non-linear dependencies in stock returns.} Russian-doll risk model, in analogy with Russian ``matryoshkas". Here one may wonder if a Russian-doll model is simply equivalent to a conventional multi-factor risk model\footnote{\, For a partial list of works on factor models, see, {\em e.g.}, fn. \ref{foot3} and
(Bansal and Viswanathan, 1993),
(Black, 1972),
(Black {\em et al}, 1972),
(Blume and Friend, 1973),
(Brandt {\em et al}, 2010),
(Campbell, 1987),
(Campbell {\em et al}, 2001),
(Campbell and Shiller, 1988),
(Carhart, 1997), (Cochrane, 2001),
(Connor and Korajczyk, 1988, 1989),
(DeBondt and Thaler, 1985),
(Dhrymes {\em et al}, 1984),
(Fama and French, 1996),
(Fama and MacBeth, 1973),
(Ferson and Harvey, 1991, 1999),
(Hall {\em et al}, 2002),
(Jagannathan and Wang, 1996),
(Kakushadze, 2014, 2015),
(Kakushadze and Liew, 2014),
(Kothari and Shanken, 1997),
(Lehmann and Modest, 1988),
(Lintner, 1965),
(Lo, 2010),
(Lo and MacKinlay, 1990),
(MacKinlay, 1995),
(Merton, 1973),
(Mukherjee and Mishra, 2005),
(Ng {\em et al}, 1992),
(Ross, 1976),
(Schwert, 1990),
(Sharpe, 1964),
(Whitelaw, 1997),
(Zhang, 2010).}  with fewer risk factors. The answer is no. A Russian-doll risk model is actually equivalent to a conventional multi-factor risk model with more risk factors, but with mostly (or completely -- in the case where no FCM need be computed) diagonal factor covariance matrix.

{}In a Russian-doll risk model, essentially, one is modeling off-diagonal elements in SCM via factor loadings and specific risks for the risk factors arising at each intermediate step in the successive nested embedding. In this regard, it is natural to wonder, given a set of risk factors, if we wish to model their FCM via a factor model, what should be the fewer new risk factors for this nested factor model? The answer is evident in the case of industry classification based risk factors with a tree-like hierarchy such as ``sector $\rightarrow$ industry $\rightarrow$ sub-industry" in the case of BICS (Bloomberg Industry Classification System) -- other industry classifications such as GICS, ICB, SIC, NAICS, {\em etc.}, use other namings for the levels in their hierarchical trees, and the number of levels in such trees need not be 3 either. Using the BICS example to illustrate our idea here, if we start with a relatively large number $K$ of sub-industries, the risk factors for modeling FCM for sub-industries via a factor model would be industries, and the risk factors for modeling FCM for industries via a factor model would be sectors. The number of BICS sectors is only 10. If need be, we can take this nested Russian-doll embedding a step further and use the ``market" ({\em e.g.}, equally weighted average of all stock return, or the intercept in the regression terminology) as the sole risk factor for modeling FCM for sectors via a single-factor model, thereby eliminating the necessity of computing non-diagonal FCM altogether: $1 \times 1$ FCM for the remaining single factor is just its variance.

{}In the case of binary industry classification based risk factors the Russian-doll embedding, as we saw above, is natural. What about non-binary style risk factors? In this case there is no simple prescription for reducing their number. However, in practical applications there is no need to reduce the number of style risk factors as this number is already small, especially for short-horizon returns -- recently it was argued in (Kakushadze, 2014) that the number of relevant style risk factors for overnight (and similarly short-horizon) returns is at most 4. As we discuss in detail in Section \ref{sec3}, this allows us to construct a Russian-doll model for a combination of style plus industry classification based risk factors, where we keep the style risk factors intact and reduce the number of industry factors via a Russian-doll embedding. At the end, we only need to compute FCM for a small number of risk factors, which include the original style risk factors (plus, {\em e.g.}, sector or ``market" risk factors).

{}One question that arises in all multi-factor model building is how to compute FCM and specific risk in a consistent fashion. There exist nontrivial algorithms for doing this, and they are typically deemed proprietary, so they are not discussed in the literature. In Section 3 we explain why such algorithms are needed and why a naive approach fails. In the context of Russian-doll risk models one can use such an algorithm at each stage of modeling FCM by a factor model. However, in Section 4 we discuss an illustrative example and a cruder way of constructing a Russian-doll risk model in the case of industry classification based risk factors without the need to employ such sophisticated proprietary algorithms. While this is admittedly a ``layman's" way of building a Russian-doll risk model, it serves the purpose of illustrating the construction and enables us to run backtests on some simple intraday mean-reversion alphas to make sure that it adds value.

\section{Multi-factor Risk Models}\label{sec2}

{}In a multi-factor risk model, a sample covariance matrix $C_{ij}$ for $N$ stocks,\footnote{\, The sample covariance matrix is computed based on a time series of stock returns, {\em e.g.}, daily close-to-close returns.} $i,j = 1,\dots,N$  is modeled by $\Gamma_{ij}$ given by
\begin{eqnarray}\label{Gamma}
 &&\Gamma \equiv \Xi + \Omega~\Phi~\Omega^T\\
 && \Xi_{ij} \equiv \xi_i^2 ~\delta_{ij}
\end{eqnarray}
where $\delta_{ij}$ is the Kronecker delta; $\Gamma_{ij}$ is an $N\times N$ matrix; $\xi_i$ is specific risk (a.k.a. idiosyncratic risk) for each stock; $\Omega_{iA}$ is an $N\times K$ factor loadings matrix; and $\Phi_{AB}$ is a $K\times K$ factor covariance matrix, $A,B=1,\dots,K$. {\em I.e.}, the random processes $\Upsilon_i$ corresponding to $N$ stock returns are modeled via $N$ random processes $\chi_i$ (corresponding to specific risk) together with $K$ random processes $f_A$ (corresponding to factor risk):
\begin{eqnarray}\label{Upsilon}
 &&\Upsilon_i = \chi_i + \sum_{A=1}^K \Omega_{iA}~f_A\\
 &&\left<\chi_i, \chi_j\right> = \Xi_{ij}\\
 &&\left<\chi_i, f_A\right> = 0\\
 &&\left<f_A, f_B\right> = \Phi_{AB}\\
 &&\left<\Upsilon_i, \Upsilon_j\right> = \Gamma_{ij}
\end{eqnarray}
The main reason for replacing the sample covariance matrix $C_{ij}$ by $\Gamma_{ij}$ is that the off-diagonal elements of $C_{ij}$ typically are not expected to be too stable out-of-sample. A constructed factor model covariance matrix $\Gamma_{ij}$ is expected to be much more stable as the number of risk factors, for which the factor covariance matrix $\Phi_{AB}$ needs to be computed, is $K\ll N$. Also, if $M<N$, where $M+1$ is the number of observations in each time series, then $C_{ij}$ is singular with $M$ nonzero eigenvalues. Assuming all $\xi_i>0$ and $\Phi_{AB}$ is positive-definite, then $\Gamma_{ij}$ is automatically positive-definite (and invertible).

\subsection{Out-of-Sample Stability}

{}While the factor model covariance matrix $\Gamma_{ij}$ is expected to be more stable than the sample covariance matrix $C_{ij}$, in practical applications, if the number of risk factors $K$ is too large, the factor covariance matrix $\Phi_{AB}$ -- and consequently $\Gamma_{ij}$ -- may not be stable enough. In fact, if $M < K$, then $\Phi_{AB}$ itself would be singular. This is the case not only when the number of available observations in the time series of stock returns is limited, but also when it is not desirable to consider long lookbacks, {\em e.g.}, when the risk model is intended to be used for (ultra-)short horizon strategies. In such cases, due to the ephemeral nature of underlying alphas, often it makes little to no sense to go back years or even months when computing the factor covariance matrix and specific risk. This then implies that the number of risk factors $K$ cannot be too large. However, in some cases there is a way to effectively enlarge the number of risk factors by capturing a partial (and out-of-sample stable) effects of more than $M$ risk factors. We discuss this methodology in the next section, first for the case of a binary industry classification, and then for a more general setting.

{}However, before we do this, let us comment on the principal component approach to multi-factor risk models. What if we simply take the first $K$ principal components of the sample covariance matrix as our risk factors?\footnote{\, This is essentially the APT (Arbitrage Pricing Theory) approach; for a partial list, see, {\em e.g.},
(Bansal and Viswanathan, 1993),
(Burmeister and Wall, 1986),
(Chen {\em et al}, 1986),
(Connor and Korajczyk, 1988, 1989),
(Dhrymes {\em et al}, 1984),
(Lehmann and Modest, 1988),
(Ross, 1976).} Here we run into two issues. First, only $M$ eigenvalues of $C_{ij}$ are nonzero, so $K\leq M$ in this approach, and if $M$ is small, we are back to where we started. Second, since the principal components are based on off-diagonal components of $C_{ij}$, typically, they are inherently unstable out-of-sample. In contrast, in the Russian-doll risk models the main idea is to dramatically reduce or altogether eliminate the risk factors
for which a sample factor covariance matrix needs to be computed, thereby reducing or eliminating such out-of-sample instability. Furthermore, the sample factor covariance matrix of industry-based risk factors typically is much more stable out-of-sample than the stock sample covariance matrix or any of its derivatives, such as its principal components.\footnote{\, This is one of the main reasons why multi-factor risk models based on constructed risk factors, such as BARRA, Northfield, Axioma, {\em etc.}, are more popular and more widely used in quantitative trading than those based on principal components (albeit traders without the know-how for custom-building constructed risk models sometimes opt for principal components).}

\section{Nested Russian-Doll Risk Models}\label{sec3}

{}The general idea behind nested risk models is simple. Suppose we have $K$ risk factors $f_A$. If the desirable/available number of observations $M < K$, then the factor covariance matrix $\Phi_{AB}$ is singular. Even if $M \geq K$, unless $K\ll M$, $\Phi_{AB}$ typically is not expected to be too stable out-of-sample. So, the idea is to model $\Phi_{AB}$ itself via a factor model (as opposed to computing it as a sample covariance matrix of the risk factors $f_A$):
\begin{equation}
 \Phi_{AB} = \zeta_A^2~\delta_{AB} + \sum_{a,b=1}^F\Lambda_{Aa}~\Psi_{ab}~\Lambda_{Bb}
\end{equation}
where $\zeta_A$ is the specific risk for $f_A$; $\Lambda_{Aa}$, $A=1,\dots,K$, $a=1,\dots,F$ is the corresponding factor loadings matrix; and $\Psi_{ab}$ is the factor covariance matrix for the underlying risk factors $g_a$, $a=1,\dots,F$, where we assume that $F\ll K$.

{}With the factor covariance matrix $\Phi_{AB}$ modeled as above, we have the following factor model covariance matrix for stocks:
\begin{equation}\label{Gamma.doll}
 \Gamma_{ij} = \xi_i^2~\delta_{ij} + \sum_{A=1}^K \zeta_A^2~\Omega_{iA}~\Omega_{jA} + \sum_{a,b=1}^F {\widetilde \Omega}_{ia}~\Psi_{ab}~{\widetilde \Omega}_{jb}
\end{equation}
where (in matrix notation)
\begin{equation}
 {\widetilde \Omega} \equiv \Omega~\Lambda
\end{equation}
Note that the first and third terms on the r.h.s. in (\ref{Gamma.doll}) comprise nothing but an $F$-factor model. However, it is the presence of the second term that makes a difference. In addition to the $F$ risk factors (the third term on the r.h.s. in (\ref{Gamma.doll})), it models off-diagonal terms in $\Gamma_{ij}$ via the factor loadings $\Omega_{iA}$ and the specific risks $\zeta_A$ for the factors $f_A$. If computed properly (see below), specific risk -- just as total risk -- is much more stable out-of-sample than sample correlations. This is because -- just as for total risk -- specific risk corresponds to variances (as opposed to off-diagonal elements in a sample covariance matrix). This makes the nested ``Russian-doll" (``matryoshka") risk model construction (\ref{Gamma.doll}) much more stable out-of-sample than the direct construction (\ref{Gamma}), yet it captures off-diagonal contributions in $\Gamma_{ij}$ beyond what an $F$-factor model would account for.

{}In fact, (\ref{Gamma.doll}) is a $(K+F)$-factor model of a special form. Indeed, we can rewrite (\ref{Gamma.doll}) as follows:
\begin{equation}
 \Gamma = \Xi + \omega~\phi~\omega^T
\end{equation}
where $\omega$ is an $N\times (K+F)$ factor loadings matrix of the form
\begin{eqnarray}
 &&\omega_{iA} = \Omega_{iA}\\
 &&\omega_{ia} = {\widetilde\Omega}_{ia}
\end{eqnarray}
and $\phi$ is a $(K+F)\times(K+F)$ factor covariance matrix of the form
\begin{eqnarray}
 &&\phi_{AB} = \zeta_A^2~\delta_{AB}\\
 &&\phi_{ab} = \Psi_{ab}\\
 &&\phi_{Aa} = \phi_{aA} = 0
\end{eqnarray}
So $\phi$ is almost diagonal -- except for the off-diagonal elements in $\Psi_{ab}$.

\subsection{Binary Industry Classification}\label{sub3.1}

{}Our discussion above might sound like a free lunch. It is not. There is still work to be done. In particular, it is not always evident what the risk factors $g_a$ for modeling the risk factors $f_A$ should be. Fortunately, there are cases where (most of) the required work has already been done. Binary industry classifications are one such case. First we keep our discussion general and then apply the binary property.

{}For concreteness we will use the BICS terminology for the levels in the industry classification, albeit this is not critical here. Also, BICS has three levels ``sector $\rightarrow$ industry $\rightarrow$ sub-industry" (where ``sub-industry" is the most detailed level). For definiteness, we will assume three levels here, albeit generalization to more levels is straightforward. So, we have: $N$ stocks labeled by $i=1,\dots,N$; $K$ sub-industries labeled by $A=1,\dots,K$; $F$ industries labeled by $a=1,\dots,F$; and $L$ sectors labeled by $\alpha=1,\dots,L$. A nested Russian-doll risk model then is constructed as follows:
\begin{eqnarray}
 &&\Gamma_{ij} = \xi_i^2~\delta_{ij} + \sum_{A,B=1}^K \Omega_{iA}~\Phi_{AB}~\Omega_{jB}\\
 &&\Phi_{AB} = \zeta_A^2~\delta_{AB} + \sum_{a,b=1}^F\Lambda_{Aa}~\Psi_{ab}~\Lambda_{Bb}\\
 &&\Psi_{ab} = \eta_a^2~\delta_{ab} + \sum_{\alpha,\beta=1}^L\Delta_{a\alpha}~\Theta_{\alpha\beta}~\Delta_{b\beta}\label{Psi}\\
 &&\Gamma_{ij} = \xi_i^2~\delta_{ij} + \sum_{A=1}^K \zeta_A^2~\Omega_{iA}~\Omega_{jA} + \sum_{a=1}^F \eta_a^2~{\widetilde \Omega}_{ia}~{\widetilde \Omega}_{ja} + \sum_{\alpha,\beta=1}^L {\widehat \Omega}_{i\alpha}~\Theta_{\alpha\beta}~{\widehat \Omega}_{j\beta}\label{Gamma.doll.sec}
\end{eqnarray}
where\footnote{\, Here $\eta_a$ is the specific risk for the risk factors $g_a$ corresponding to industries, $\Theta_{\alpha\beta}$ is the factor covariance matrix for sectors, and $\Delta_{a\alpha}$ is the corresponding factor loadings matrix. Other notations are as above and self-explanatory.}
\begin{eqnarray}
 && {\widetilde \Omega} \equiv \Omega~\Lambda\\
 && {\widehat \Omega} \equiv {\widetilde\Omega}~\Delta
\end{eqnarray}
Note that it is the second and third terms on the r.h.s. of (\ref{Gamma.doll.3}) that make (in an out-of-sample stable fashion) this construction different from an $L$-factor model corresponding to sectors as risk factors.

{}Here too, we can view (\ref{Gamma.doll.3}) as a larger, $(K+F+L)$-factor model of a special form:
\begin{equation}
 \Gamma = \Xi + \omega~\phi~\omega^T
\end{equation}
where $\omega$ is an $N\times (K+F+L)$ factor loadings matrix of the form
\begin{eqnarray}
 &&\omega_{iA} = \Omega_{iA}\\
 &&\omega_{ia} = {\widetilde\Omega}_{ia}\\
 &&\omega_{i\alpha} = {\widehat\Omega}_{i\alpha}
\end{eqnarray}
and $\phi$ is a $(K+F+L)\times(K+F+L)$ factor covariance matrix of the form
\begin{eqnarray}
 &&\phi_{AB} = \zeta_A^2~\delta_{AB}\\
 &&\phi_{ab} = \eta_a^2~\delta_{ab}\\
 &&\phi_{\alpha\beta} = \Theta_{\alpha\beta}\\
 &&\phi_{Aa} = \phi_{aA} = \phi_{A\alpha} = \phi_{\alpha A} = \phi_{a\alpha} = \phi_{\alpha a} = 0
\end{eqnarray}
So $\phi$ is almost diagonal -- except for the off-diagonal elements in $\Theta_{\alpha\beta}$.

\subsection{``Single-Factor" Russian-Doll Risk Model}

{}We can take the above construction one step further and reduce it to a ``single-factor" model by modeling $\Theta_{\alpha\beta}$ via a 1-factor risk model. The factor loadings matrix $\Pi_{\alpha}$ is just a column, an ($L\times 1$) matrix, which can be chosen to be simply the intercept: $\Pi_\alpha\equiv 1$. The corresponding factor covariance matrix $X$ is just a positive number ($1\times 1$ matrix), so we have
\begin{equation}
 \Theta_{\alpha\beta} = \sigma_{\alpha}^2~\delta_{\alpha\beta} + X
\end{equation}
and
\begin{equation}\label{Gamma.doll.3}
 \Gamma_{ij} = \xi_i^2~\delta_{ij} + \sum_{A=1}^K \zeta_A^2~\Omega_{iA}~\Omega_{jA} + \sum_{a=1}^F \eta_a^2~{\widetilde \Omega}_{ia}~{\widetilde \Omega}_{ja} +
 \sum_{\alpha=1}^L \sigma_\alpha^2~{\widehat \Omega}_{i\alpha}~{\widehat \Omega}_{j\alpha} + X~{\overline\Omega}_i~{\overline \Omega}_j
\end{equation}
where
\begin{equation}
 {\overline\Omega}_i \equiv \sum_{\alpha=1}^L {\widehat\Omega}_{i\alpha}~\Pi_\alpha = \sum_{\alpha=1}^L {\widehat\Omega}_{i\alpha}
\end{equation}
This ``single-factor" model is actually a $(K+F+L+1)$-factor model with the factor loadings matrix given by $\omega = (\Omega, {\widetilde \Omega}, {\widehat \Omega}, {\overline \Omega})$ and a diagonal factor covariance matrix given by $\phi = \mbox{diag}\left(\zeta_A^2~\delta_{AB}, \eta_a^2~\delta_{ab}, \sigma_\alpha^2~\delta_{\alpha\beta}, X\right)$.

{}Further, note that if we set $X=0$, we obtain a ``zero-factor" Russian-doll model, where all off-diagonal elements are modeled via the second, third and forth terms on the r.h.s. of (\ref{Gamma.doll.3}), which is actually a $(K+F+L)$-factor model with a diagonal factor covariance matrix.

\subsection{Binary Property}

{}The binary property implies that each stock belongs to one and only one sub-industry, industry and sector. The factor loadings matrices $\Omega_{iA}$, $\Lambda_{Aa}$ and $\Delta_{a\alpha}$ are given by
\begin{eqnarray}
 &&\Omega_{iA} = \delta_{G(i), A}\\
 &&\Lambda_{Aa} = \delta_{S(A), a}\\
 &&\Delta_{a\alpha} = \delta_{T(a), \alpha}
\end{eqnarray}
where $G$ is the map between stocks and sub-industries, $S$ is the map between sub-industries and industries, and $T$ is the map between industries and sectors: \begin{eqnarray}
 &&G:\{1,\dots,N\}\mapsto\{1,\dots,K\}\\
 &&S:\{1,\dots,K\}\mapsto\{1,\dots,F\}\\
 &&T:\{1,\dots,F\}\mapsto\{1,\dots,L\}
\end{eqnarray}
This implies that
\begin{eqnarray}
 &&{\widetilde\Omega}_{ia} = \delta_{{\widetilde G}(i), a}\\
 &&{\widehat\Omega}_{i\alpha} = \delta_{{\widehat G}(i), \alpha}
\end{eqnarray}
where ${\widetilde G}\equiv SG$ is the map between stocks and industries, and ${\widehat G}\equiv T{\widetilde G} = TSG$ is the map between stocks and sectors. Eq. (\ref{Gamma.doll.sec}) then simplifies as follows:
\begin{equation}
 \Gamma_{ij} = \xi_i^2~\delta_{ij} + \zeta_{G(i)}^2~\delta_{G(i),G(j)} + \eta_{{\widetilde G}(i)}^2~\delta_{{\widetilde G}(i),{\widetilde G}(j)} + \Theta_{{\widehat G}(i),{\widehat G}(j)}
\end{equation}
The key simplifying feature of a binary industry classification\footnote{\, See footnote \ref{foot.binary} below.} is that the risk factors $f_A$, $g_a$ and $h_\alpha$ (where $h_\alpha$ are the risk factors for the factor model (\ref{Psi}) for $\Psi_{ab}$) are explicitly known once the industry classification tree is specified: $f_A$ correspond to the sub-industry risk, $g_a$ correspond to the industry risk, and $h_\alpha$ correspond to the sector risk. Therefore, constructing the Russian-doll risk model boils down to calculating $\Theta_{\alpha\beta}$ factor covariance matrix for the sector risk factors $h_{\alpha}$ and also fixing the specific risks $\xi_i$, $\zeta_A$ and $\eta_a$ (see below).

\subsection{Non-Binary Generalization}

{}The beauty of dealing with a binary\footnote{\, Here one can also work with ``non-binary" industry classifications in the sense that each ticker may belong to, say, more than one sub-industries, {\em e.g.}, in the case of conglomerates. However, typically the number of such tickers is relatively small and the number of sub-industries such a ticker belongs to typically is a few to several. Here, not to muddy the waters, we will stick to binary industry classifications. We will deal with non-binary style risk factors instead.\label{foot.binary}} classification is that the hierarchy of risk factors ({\em i.e.}, $f_A \leftarrow g_a \leftarrow h_\alpha$) is fixed by the classification hierarchy ({\em i.e.}, ``sector $\rightarrow$ industry $\rightarrow$ sub-industry"), and the latter is readily available -- the industry classification provider has already done all the hard work of analyzing companies' products and/or services, revenue sources, {\em etc.}, that determine the company taxonomy and industry classification.\footnote{\, While there are several commercially available industry classifications with their own proprietary methodologies, the top performing ones are relatively similar. Here we do not suggest that one use any particular industry classification -- it is a matter of preference and access.} With non-industry risk factors it is not always as straightforward to identify a nested hierarchy of risk factors. {\em E.g.}, in the case of principal component based risk factors there is no evident guiding principle to do so.

{}However, not all is lost. In practice, the most popular multi-factor risk models combine non-binary style risk factors and binary industry risk factors.\footnote{\, Some models use quasi-binary industry risk factors with conglomerates assigned fractional weights in several different industries -- as mentioned above, here for the sake of simplicity we stick to binary industry risk factors.} The number of style factors typically is substantially smaller than the number of industry factors, especially for (ultra-)short horizon models. {\em E.g.}, recently in (Kakushadze, 2014) it was argued that for overnight returns there are essentially 4 relevant style risk factors. The question we wish to address here is whether we can start with a few style risk factors plus many more industry based risk factors (typically, $\sim 100$ or more), and build a Russian-doll risk model.

{}It is precisely the fact that we have only a few style risk factors that allows us to build a Russian-doll model -- this is because there is no need to reduce the number of style risk factors, only that of the industry based risk factors. So, the idea here is quite simple. We will use the mid-Greek symbols $\mu, \nu,\dots$ to label the style risk factors, and we use the $i, A, a, \alpha$ labels as above. Let ${\widetilde A} \equiv (A, \mu)$, ${\widetilde a} \equiv (a, \mu)$ and ${\widetilde \alpha} \equiv (\alpha, \mu)$. Let $U$ be the number of style risk factors. Let ${\widetilde K}\equiv K + U$, ${\widetilde F}\equiv F+U$ and ${\widetilde L}\equiv L+U$. Then we can apply the method of Subsection \ref{sub3.1} to $i, {\widetilde A}, {\widetilde a}, {\widetilde \alpha}$:
\begin{eqnarray}
 &&\Gamma_{ij} = \xi_i^2~\delta_{ij} + \sum_{{\widetilde A},{\widetilde B}=1}^{\widetilde K} \Omega_{i{\widetilde A}}~\Phi_{{\widetilde A}{\widetilde B}}~\Omega_{j{\widetilde B}}\\
 &&\Phi_{{\widetilde A}{\widetilde B}} = \zeta_{\widetilde A}^2~\delta_{{\widetilde A}{\widetilde B}} + \sum_{{\widetilde a},{\widetilde b}=1}^{\widetilde F}\Lambda_{{\widetilde A}{\widetilde a}}~\Psi_{{\widetilde a}{\widetilde b}}~\Lambda_{{\widetilde B}{\widetilde b}}\\
 &&\Psi_{{\widetilde a}{\widetilde b}} = \eta_{\widetilde a}^2~\delta_{{\widetilde a}{\widetilde b}} + \sum_{{\widetilde \alpha},{\widetilde \beta}=1}^{\widetilde L}\Delta_{{\widetilde a}{\widetilde \alpha}}~\Theta_{{\widetilde \alpha}{\widetilde \beta}}~\Delta_{{\widetilde b}{\widetilde \beta}}\label{Psi.nb}
\end{eqnarray}
So we have
\begin{equation}
 \Gamma_{ij} = \xi_i^2~\delta_{ij} + \sum_{{\widetilde A}=1}^{\widetilde K} \zeta_{\widetilde A}^2~\Omega_{i{\widetilde A}}~\Omega_{j{\widetilde A}} + \sum_{{\widetilde a}=1}^{\widetilde F} \eta_{\widetilde a}^2~{\widetilde \Omega}_{i{\widetilde a}}~{\widetilde \Omega}_{j{\widetilde a}} + \sum_{{\widetilde \alpha},{\widetilde \beta}=1}^{\widetilde L} {\widehat \Omega}_{i{\widetilde \alpha}}~\Theta_{{\widetilde \alpha}{\widetilde \beta}}~{\widehat \Omega}_{j{\widetilde \beta}}\label{Gamma.doll.sec.nb}
\end{equation}
where
\begin{eqnarray}
 &&{\widetilde \Omega}_{i{\widetilde a}} = \sum_{{\widetilde A}=1}^{\widetilde K} \Omega_{i{\widetilde A}}~\Lambda_{{\widetilde A}{\widetilde a}}\\
 &&{\widehat \Omega}_{i{\widetilde \alpha}} = \sum_{{\widetilde a}=1}^{\widetilde F} {\widetilde \Omega}_{i{\widetilde a}}~\Delta_{{\widetilde a}{\widetilde \alpha}}
\end{eqnarray}
and we further have
\begin{eqnarray}
 &&\Lambda_{{\widetilde A}{\widetilde a}} = \mbox{diag}\left(\Lambda_{Aa}, \delta_{\mu\nu}\right)\\
 &&\Delta_{{\widetilde a}{\widetilde \alpha}} = \mbox{diag}\left(\Delta_{a\alpha}, \delta_{\mu\nu}\right)\\
 &&\zeta_\mu = 0\\
 &&\eta_\mu = 0\\
 &&\Phi_{\mu\nu} = \Psi_{\mu\nu} = \Theta_{\mu\nu}\\
 &&\Phi_{A\mu} = \sum_{a=1}^F \Lambda_{Aa}~\Psi_{a\mu}\\
 &&\Psi_{a\mu} = \sum_{\alpha=1}^L \Delta_{a\alpha}~\Theta_{\alpha\mu}
\end{eqnarray}
So, at the end, everything is fixed via the ${\widetilde L}\times {\widetilde L}$ factor covariance matrix $\Theta_{{\widetilde\alpha}{\widetilde\beta}}$ and the specific risks $\xi_i$, $\zeta_A$ and $\eta_a$. As before, the factor loadings matrices $\Omega_{iA}$, $\Lambda_{Aa}$ and $\Delta_{a\alpha}$ are binary.

\subsection{Fixing Factor Covariance Matrix and Specific Risk}\label{fix.cov.mat}

{}One may ask, this is all good, but how do I compute the remaining factor covariance matrix $\Theta$ and the specific risks $\xi_i$, $\zeta_A$ and $\eta_a$? A simple answer is that, if one knows how to compute the factor covariance matrix and specific risk for the usual factor model (\ref{Gamma}), then the same methods can be applied to the Russian-doll factor models with some straightforward adjustments. However, the methodologies for computing the factor covariance and specific risk are usually deemed proprietary\footnote{\, And so do the author and Quantigic$^\circledR$ Solutions LLC.} and, therefore, are well outside of the scope of this note. Nonetheless, here we wish to discuss what appears to be a common misconception for how to compute the factor covariance matrix and specific risk and point out where and why it fails.

{}This misconception apparently stems from a formal similarity between (\ref{Upsilon}) and a linear (cross-sectional) regression
\begin{equation}
 R_{is} = \epsilon_{is} + \sum_{A=1}^K \beta_{iAs}~f_{As}
\end{equation}
where $s$ labels time series, $R_{is}$ are stock returns ({\em e.g.}, daily close-to-close returns, in which case $s$ labels trading dates), $\epsilon_{is}$ are the regression residuals (for each date $s$), $\beta_{iAs}$ are factor betas, and $f_{As}$ are factor returns (note that we have $K$ factors). If for some period $\beta_{iAs}$ are independent of $s$, $\beta_{iAs} \equiv \beta_{iA}$ ({\em e.g.}, we compute them monthly), then we can identify $\beta_{iA}$ with $\Omega_{iA}$, and, for each date $s$, $R_{is}$ is identified with $\Upsilon_i$, $\epsilon_{is}$ is identified with $\chi_i$, and $f_{As}$ is identified with $f_A$. It is then tempting to erroneously conclude that the factor covariance matrix $\Phi_{AB}$ is simply given by $\left\langle f_A,f_B\right\rangle$, while the specific variance $\xi_i^2$ is given by $\left\langle \epsilon_i,\epsilon_i\right\rangle$, where the covariance $\left\langle *,*\right\rangle$ is computed over the time series (and we have suppressed the index $s$). However, a quick computation reveals the fallacy of this approach. Indeed, from the definition of the linear regression (without intercept and unit weights) we have (in matrix notation)
\begin{eqnarray}
 &&f = \left(\Omega^T~\Omega\right)^{-1}\Omega^T~R\\
 &&\epsilon = \left[1 - Q\right] R
\end{eqnarray}
where (note that $Q = Q^T$ is a projection operator: $Q^2 = Q$)
\begin{equation}
 Q \equiv \Omega\left(\Omega^T~\Omega\right)^{-1}\Omega^T
\end{equation}
Consequently, we have:
\begin{eqnarray}
 &&\left\langle\epsilon,\epsilon^T\right\rangle = \left[1 - Q\right] C \left[1 - Q\right]\\
 &&\Omega\left \langle f, f^T\right\rangle\Omega^T = Q~C~Q
\end{eqnarray}
where $C_{ij}\equiv \left\langle R_i, R_j\right\rangle$ is the sample covariance matrix. Now we can immediately see the issue with identifying $\Phi_{AB}$ with $\left\langle f_A,f_B\right\rangle$ and $\xi_i^2$ with $\left\langle \epsilon_i,\epsilon_i\right\rangle$. The total variance $\Gamma_{ii}$ according to the factor model is given by
\begin{equation}
 \Gamma_{ii} = \xi_i^2 + \sum_{A,B=1}^K \Omega_{iA}~\Phi_{AB}~\Omega_{iB}
\end{equation}
With the above (erroneous) identifications, the factor model total variance $\Gamma_{ii}$ does {\em not} coincide\footnote{\, Only the trace coincides: $\mbox{Tr}\left(\left\langle\epsilon,\epsilon^T\right\rangle+\Omega\left \langle f, f^T\right\rangle\Omega^T\right) = \mbox{Tr}(C)$.} with the in-sample total variance $C_{ii}$ -- and a factor model had better reproduce the in-sample total variance (while attempting to predict out-of-sample total variance as precisely as possible). Also note that if we keep the above identification of $\Phi_{AB}$ with $\left\langle f_A,f_B\right\rangle$ and simply {\em define} $\xi_i^2\equiv C_{ii} - \sum_{A,B=1}^K \Omega_{iA}~\Phi_{AB}~\Omega_{iB}$, generally we will (unacceptably) have some negative $\xi_i^2$.

\section{An Illustrative Example}\label{sec4}

{}While computing the factor covariance matrix and specific risk(s) is nontrivial (and a proprietary topic), the Russian-doll risk modeling allows to bypass such complications by using simple heuristics. We emphasize that using the full-fledged risk modeling by carefully computing the factor covariance matrix and specific risk(s) generally yields better results. However, if the latter is not possible, the heuristic approach, which we illustrate in this section, provides an approximate method for incorporating off-diagonal correlations into the covariance matrix.

{}The idea here is very simple. To avoid the headaches discussed in Subsection \ref{fix.cov.mat}, let us simply avoid computing {\em any} factor covariance matrix. We are then led to consider the ``single-factor" Russian-doll factor model (\ref{Gamma.doll.3}), where the only ``factor covariance matrix" is the $1\times 1$ matrix $X$, which is in fact the {\em variance} of the single factor, which in turn can be interpreted as the overall ``market" exposure.\footnote{\, Here the ``market" return is defined as equally weighted average of all stock returns in the universe labeled by $i\in\{1,\dots,N\}$.} Not to overcomplicate our discussion here, let us stick to the binary case. Then, using the same notations as above, we have:
\begin{equation}
 \Gamma_{ij} = \xi_i^2~\delta_{ij} + \zeta_{G(i)}^2~\delta_{G(i),G(j)} + \eta_{{\widetilde G}(i)}^2~\delta_{{\widetilde G}(i),{\widetilde G}(j)} + \sigma_{{\widehat G}(i)}^2~\delta_{{\widehat G}(i),{\widehat G}(j)} + X
\end{equation}
So, the total variance is given by
\begin{equation}
 \Gamma_{ii} = \xi_i^2 + \zeta_{G(i)}^2 + \eta_{{\widetilde G}(i)}^2 + \sigma_{{\widehat G}(i)}^2 + X
\end{equation}
As above, we wish to identify $\Gamma_{ii}$ with the in-sample total variance $C_{ii}$. This gives us $N$ equations for $N+K+F+L+1$ unknowns $\xi_i$, $\zeta_A$, $\eta_a$, $\sigma_\alpha$ and $X$. Then, as before, the main issue here is that generally some $\xi_i^2$, $\zeta_A^2$, $\eta_a^2$, $\sigma_\alpha^2$ and/or $X$
will (unacceptably) be negative. Thus, if we require that all $\xi_i^2$, $\zeta_A^2$, $\eta_a^2$, $\sigma_\alpha^2$ and $X$ are non-negative, then we get $\zeta_A^2 \leq \mbox{min}(C_{ii})$, $\eta_a^2\leq \mbox{min}(C_{ii})$, $\sigma_\alpha^2\leq \mbox{min}(C_{ii})$ and $X\leq \mbox{min}(C_{ii})$, and since the variances $C_{ii}$ have a skewed (theoretically, log-normal) distribution, this implies that $\zeta_A$, $\eta_a$, $\sigma_\alpha$ and $X$ will have a small effect on most tickers with larger $C_{ii}$, including on the corresponding off-diagonal elements, {\em i.e.}, the correlations involving such tickers will be small. Here we discuss a simple heuristic ``fix" (or ``hack").

{}The key observation here is that, if $C_{ii}$ were more uniform, then requiring that all $\xi_i^2$, $\zeta_A^2$, $\eta_a^2$, $\sigma_\alpha^2$ and $X$ are non-negative generically would not yield small correlations. Therefore, let us factor out the non-uniformity in $C_{ii}$ by considering the correlation matrix $\Psi_{ij}$ instead of $C_{ij}$:
\begin{equation}
 C_{ij} \equiv \sqrt{C_{ii}}\sqrt{C_{jj}}~\Psi_{ij}
\end{equation}
where $\Psi_{ii} = 1$. So, {\em ad hoc}, instead of $C_{ij}$, we now model $\Psi_{ij}$ via a Russian-doll factor model, {\em i.e.}, we identify $\Gamma_{ii}$ with $\Psi_{ii}$, so we have $\Gamma_{ii} = 1$ and
\begin{equation}
 \xi_i^2 + \zeta_{G(i)}^2 + \eta_{{\widetilde G}(i)}^2 + \sigma_{{\widehat G}(i)}^2 + X = 1
\end{equation}
and we are left with only $K+F+L+1$ unknowns $\zeta_A$, $\eta_a$, $\sigma_\alpha$ and $X$.

{}To make progress, let us observe that there is no unique solution or magic prescription here.\footnote{\, Even in the aforesaid proprietary algorithms one must make certain (sophisticated) choices.} With this in mind, let us consider the following simple Ansatz:
\begin{equation}\label{crude}
 \xi_i^2 =\zeta^2_A = \eta_a^2 = \sigma_\alpha^2 = X = 1/5
\end{equation}
{\em I.e.}, the ``market", sectors, industries and sub-industries are assumed to contribute into the total variance with equal weights, same as the stock-specific (idiosyncratic) risk. Again, this is a simplified assumption, but it will suffice for {\em illustrative} purposes.

\subsection{Horse Race}

{}Next, we wish to see if the above simplified Russian-doll model adds value. One way to test this is to run a horse race given a trading universe and the corresponding expected returns (see below). On the one hand, to obtain desired holdings, we can use the Russian-doll model in optimization via Sharpe ratio maximization subject to the dollar neutrality constraint. On the other hand, we can run the same optimization with a diagonal sample covariance matrix $\mbox{diag}(C_{ii})$ subject to the dollar neutrality constraint, or even sector, industry and sub-industry neutrality constraints. In fact, optimization with a diagonal covariance matrix and subject to linear homogeneous constraints is equivalent to weighted cross-sectional regression with the columns of the loadings matrix (over which the returns are regressed) identified with the vectors of constraint coefficients and the regression weights identified with inverse variances $1/C_{ii}$ -- see (Kakushadze, 2015) for details. For this reason, for terminological convenience, we will refer to the horse race as between optimization (using the Russian-doll model) and weighted regression (using various constraints).\footnote{\, The remainder of this section overlaps with Section 7 of (Kakushadze, 2015) as backtesting models are similar (albeit not identical).}

\subsubsection{Notations}

{}Let us set up our notations. $P_i$, $i=1,\dots,N$ is the stock price for the stock labeled by $i$. In fact, the price for each stock is a time-series: $P_{is}$, $s=0,1,\dots,M$, where the index $s$ labels trading dates, with $s=0$ corresponding to the most recent date in the time series. We will use superscripts $O$ and $C$ (unadjusted open and close prices) and $AO$ and $AC$ (open and close prices fully adjusted for splits and dividends), so, {\em e.g.}, $P^C_{is}$ is the unadjusted close price. $V_{is}$ is the unadjusted daily volume (in shares). Also, we define the overnight return as the close-to-next-open return:
\begin{equation}
 R_{is} \equiv \ln\left({P^{AO}_{is} / P^{AC}_{i,s+1}}\right)
\end{equation}
Note that both prices in this definition are fully adjusted.

{}The portfolio is established at the open\footnote{\, This is a so-called ``delay-0" alpha -- $P^O_{is}$ is used in the alpha, and as the establishing fill price.} assuming fills at the open prices $P^O_{is}$, and liquidated at the close on the same day assuming fills at the close prices $P^C_{is}$, with no transaction costs or slippage -- our goal here is not to build a trading strategy, but to check if our Russian-doll factor model adds value. The P\&L for each stock is
\begin{equation}
 \Pi_{is} = H_{is}\left[{P^C_{is}\over P^O_{is}}-1\right]
\end{equation}
where $H_{is}$ are the desired {\em dollar} holdings. The shares bought plus sold ({\em i.e.}, for the establishing and liquidating trades combined) for each stock on each day are computed via $Q_{is} = 2 |H_{is}| / P^O_{is}$.

\subsubsection{Universe Selection}

{}Before we can run our simulations, we need to select our universe. We wish to keep our discussion here as simple as possible, so we select our universe based on the average daily dollar volume (ADDV) defined via
\begin{equation}\label{ADDV}
 A_{is}\equiv {1\over d} \sum_{r=1}^d V_{i, s+r}~P^C_{i, s+r}
\end{equation}
We take $d=21$ ({\em i.e.}, one month), and then take our universe to be top 2000 tickers by ADDV. However, to ensure that we do not inadvertently introduce a universe selection bias, we do not rebalance the universe daily. Instead, we rebalance monthly, every 21 trading days, to be precise. {\em I.e.}, we break our 5-year backtest period (see below) into 21-day intervals, we compute the universe using ADDV (which, in turn, is computed based on the 21-day period immediately preceding such interval), and use this universe during the entire such interval. The bias that we do have, however, is the survivorship bias. We take the data for the universe of tickers as of 9/6/2014 that have historical pricing data on http://finance.yahoo.com (accessed on 9/6/2014) for the period 8/1/2008 through 9/5/2014. We restrict this universe to include only U.S. listed common stocks and class shares (no OTCs, preferred shares, {\em etc.}) with BICS sector, industry and sub-industry assignments as of 9/6/2014. However, it does not appear that the survivorship bias is a leading effect here -- see Section 7 of (Kakushadze, 2015) for details. Also, ADDV-based universe selection is by no means optimal and is chosen here for the sake of simplicity. In practical applications, the trading universe of liquid stocks is carefully selected based on market cap, liquidity (ADDV), price and other (proprietary) criteria.

\subsubsection{Backtesting}

{}We run our simulation over a period of 5 years. More precisely, $M = 252\times 5$, and $s=0$ is 9/5/2014 (see above). The annualized return-on-capital (ROC) is computed as average daily P\&L divided by the (intraday -- see below) investment level $I$ (with no leverage) and multiplied by 252. The annualized Sharpe Ratio (SR) is computed as daily Sharpe ratio multiplied by $\sqrt{252}$. Cents-per-share (CPS) is computed as the total P\&L divided by total shares traded.

As mentioned above, we assume no transaction costs. This is because the transaction cost are expected to simply reduce the ROC of the optimization and weighted regression alphas by the same amount as the two strategies trade the exact same amount by design. Therefore, including the transaction costs would have no effect on the actual outperformance in the horse race. Since the purpose of the horse race is solely to examine the relative performance of the two alphas (and not to build a realistic trading strategy), including the transaction costs would only complicate things without any actual benefit.

\subsubsection{Weighted Regression Alphas}\label{sub.reg}

{}The constrains on the desired dollar holdings $H_{is}$ in our portfolio are of the form:
\begin{eqnarray}
 &&\mbox{dollar neutrality:}~~~\sum_{i=1}^N H_{is}= 0\\
 &&\mbox{sector neutrality:}~~~\sum_{i=1}^N {\widehat\Omega}_{i\alpha}~H_{is} = 0\\
 &&\mbox{industry neutrality:}~~~\sum_{i=1}^N {\widetilde\Omega}_{ia}~H_{is} = 0\\
 &&\mbox{sub-industry neutrality:}~~~\sum_{i=1}^N \Omega_{iA}~H_{is} = 0
\end{eqnarray}
Note that sector, industry and sub-industry neutrality automatically implies dollar neutrality as $\sum_{\alpha=1}^L {\widehat\Omega}_{i\alpha} \equiv 1$, $\sum_{a=1}^F {\widetilde\Omega}_{ia} \equiv 1$ and $\sum_{A=1}^K \Omega_{iA} \equiv 1$, {\em i.e.}, the intercept (that is, the unit $N$-vector) is subsumed in the loadings matrices ${\widehat\Omega}_{i\alpha}$, ${\widetilde\Omega}_{ia}$ and $\Omega_{iA}$ via linear combinations of their columns.

{}Next, for each date labeled by $s$, we run cross-sectional regressions of the returns $R_{is}$ over the corresponding loadings matrix, call it $Y$ (with indices suppressed), which has 4 different incarnations: i) for dollar neutrality $Y$ is an $N\times 1$ unit matrix (that is, the intercept); ii) for sector neutrality $Y$ is the $N\times L$ matrix ${\widehat\Omega}_{i\alpha}$; iii) for industry neutrality $Y$ is the $N\times F$ matrix ${\widetilde\Omega}_{ia}$; and iv) for sub-industry neutrality $Y$ is the $N\times K$ matrix $\Omega_{iA}$. Note that in the case i), the regression is simply over the intercept, while in the cases ii)-iv) the intercept is automatically included. The regression weights are given by $z_i \equiv 1/C_{ii}$. More precisely, for each date $s$ the sample variances $C_{iis}$ are computed out-of-sample as follows:
\begin{equation}
 C_{iis} \equiv \mbox{Var}\left(R_{i,(s+1)}, R_{i,(s+2)}, \dots, R_{i,(s+d)}\right)
\end{equation}
{\em I.e.}, for each date $s$ we take the overnight returns for the preceding $d$ trading days and compute the variances $C_{iis}$ based on the corresponding $d$-day time series. We take $d=21$ ({\em i.e.}, one month). However, to avoid unnecessary variations in the weights $z_i$ (as such variations could result in unnecessary overtrading), just as with the trading universe, we do not recompute $z_i$ daily but monthly, every 21 trading days, to be precise. {\em I.e.}, we break our 5-year backtest period into 21-day intervals, we compute the variances $C_{ii}$ based on the 21-day period immediately preceding such interval, and use these variances to compute the weights via $z_i = 1/C_{ii}$ during the entire such interval.

{}In each of the above 4 cases i)-iv), we compute the residuals $\varepsilon_{is}$ of the weighted regression and then the desired holdings $H_{is}$ via (we use matrix notation and suppress indices):
\begin{eqnarray}
 &&\varepsilon = R - Y~Q^{-1}~Y^T~Z~R\\
 &&Z\equiv\mbox{diag}(z_i)\\
 &&Q\equiv Y^T~Z~Y\\
 &&{\widetilde R} \equiv Z~\varepsilon\\
 &&H_{is} \equiv -{\widetilde R}_{is} ~ {I\over\sum_{j=1}^N \left|{\widetilde R}_{js}\right|}
\end{eqnarray}
where $Q^{-1}$ is the inverse of the matrix $Q$ (in the case i) it is a $1\times 1$ matrix), and we have:
\begin{eqnarray}
 &&\sum_{i=1}^N \left|H_{is}\right| = I\\
 &&\sum_{i=1}^N H_{is} = 0\label{d.n}
\end{eqnarray}
where $I$ is the total {\em intraday} dollar investment level (long plus short), which is the same for all dates $s$. Eq. (\ref{d.n}) implies that the portfolio is dollar neutral. This is because the ``regressed returns" ${\widetilde R}_{is}$ have 0 cross-sectional means, which in turn is due to the intercept either being included (the case i)), or being subsumed in the loadings matrix $Y$ (the cases ii)-iv)).

{}The results are given in Table 1 and P\&Ls for the 4 cases i)-iv) are plotted in Figure 1. In Table 2, for comparison purposes, we also give the results in the same 4 cases when the regression weights are set to 1. We denote the corresponding regression residuals via ${\widetilde \varepsilon}_{is}$, which we will also use below. Using inverse variances as regression weights clearly adds value,\footnote{\, More precisely, as usual, it improves SR and CPS at the expense of lowered ROC -- weighted regression based portfolios are closer to the maximized Sharpe ratio portfolios.} which is not surprising considering that this amounts to suppressing contributions of the higher volatility stocks into the portfolio by their sample variances -- as mentioned above, the weighted regression is the same as optimization via Sharpe ratio maximization with a diagonal sample covariance matrix subject to the corresponding linear homogeneous constrains.

\subsubsection{Optimized Alpha}

{}Next, we turn to the optimized alpha. In maximizing the Sharpe ratio, we use the approximate covariance matrix given by
\begin{eqnarray}
 &&\Theta_{ij} \equiv \sqrt{C_{ii}}\sqrt{C_{jj}}\left[\left(1-\zeta_{G(i)}^2 - \eta_{{\widetilde G}(i)}^2 - \sigma_{{\widehat G}(i)}^2 - X\right)\delta_{ij} +\right.\nonumber\\
 &&\,\,\,\,\,\,\,\left.+\zeta_{G(i)}^2~\delta_{G(i),G(j)} + \eta_{{\widetilde G}(i)}^2~\delta_{{\widetilde G}(i),{\widetilde G}(j)} + \sigma_{{\widehat G}(i)}^2~\delta_{{\widehat G}(i),{\widehat G}(j)} + X \right]
\end{eqnarray}
where the sample variances $C_{ii}$ are computed the same way as for the weighted regression alphas (based on 21-day intervals), and as above, for the sake of simplicity, we will set $\xi_i^2 = \zeta_A^2 =\eta_a^2 = \sigma_\alpha^2 = X = 1/5$.

{}For each date (we omit the index $s$) we are maximizing the Sharpe ratio subject to the dollar neutrality constraint:
\begin{eqnarray}
 &&{\cal S} \equiv {\sum_{i=1}^N H_i~E_i\over{\sqrt{\sum_{i,j=1}^N \Theta_{ij}~H_i~H_j}}} \rightarrow \mbox{max}\\
 &&\sum_{i=1}^N H_i = 0\label{d.n.opt}
\end{eqnarray}
where $E_i \equiv {\widetilde \varepsilon}_i$ are the expected returns (${\widetilde \varepsilon}_i$ are the {\em unit-weight} regression residuals of overnight returns over $\Omega_{iA}$ (sub-industries) -- see above). The solution is given by
\begin{equation}\label{H.opt}
 H_i = \gamma \left[\sum_{j = 1}^N \Theta^{-1}_{ij}~E_j - \sum_{j=1}^N \Theta^{-1}_{ij}~{{\sum_{k,l=1}^N \Theta^{-1}_{kl}~E_l}\over{\sum_{k,l = 1}^N \Theta^{-1}_{kl}}}\right]
\end{equation}
where $\Theta^{-1}$ is the inverse of $\Theta$, and the overall normalization constant $\gamma$ is fixed via the requirement that
\begin{equation}
 \sum_{i=1}^N \left|H_i\right| = I
\end{equation}
Note that (\ref{H.opt}) satisfies the dollar neutrality constraint (\ref{d.n.opt}).

{}The simulation results are given in Table 1 in the bottom row. The P\&L plot for this optimized alpha is included in Figure 1. It is evident that the (crude and {\em ad hoc}) Russian-doll model we use here adds value -- even though we did not compute the factor covariance matrix or specific risk and simply made some heuristic approximations. The optimized model in the bottom row of Table 1 has slightly better ROC, SR and CPS than the sub-industry-based regression model -- for all practical purposes their performances are at par. However, the difference is that the Russian-doll model -- albeit this particular version is crude and this is a layman's way of constructing it -- is a full risk model that predicts off-diagonal correlations, while what is used in the regression consists of only the loadings matrix and regression weights, and does not predict off-diagonal correlations. One can do substantially better than the crude Russian-doll model we used here for illustrative purposes by utilizing the proprietary algorithms for computing the factor covariance matrix and specific risk, albeit there is cost associated with such algorithms. However, for evident reasons, such algorithms are outside of the scope of this note.

{}It is also important to note that in the above illustrative example we purposefully did not include any style factors -- had we included style factors, we would have to compute the factor covariance matrix for the style factors plus at least the ``market" factor in the above construction, which would require utilizing a proprietary algorithm. In this regard, the above example is clearly watered down and is used here for illustrative purposes only. In the next section we further elaborate on this.

\section{Concluding Remarks}

{}To summarize, the basic idea and motivation behind Russian-doll risk model construction is that, in some cases it might be undesirable and/or impracticable to compute the factor covariance matrix based on (or, more generally, using) historical time series data. In such cases, one can use the Russian-doll construction to model the factor covariance matrix itself as a factor model thereby reducing the number of factors for which the factor covariance matrix must be computed. If need be, one can apply this (approximation) process successively to dramatically reduce the number of remaining risk factor for which the factor covariance matrix must be computed or eliminate them altogether. In this note we have discussed how to apply this construction in the case of (binary) industry classification based risk factors, and also when (non-binary) style factors are present. In practical applications the Russian-doll model building can be a powerful tool when a factor model is required to estimate/forecast off-diagonal correlations, but a full-fledged factor model is not available or impracticable to construct.

{}In this regard, let us go back to the crude construction we employed in Section \ref{sec4}. We can generalize this construction to an arbitrary risk model, not necessarily involving the Russian-doll construction. Suppose we identify our factor loadings matrix $\Omega_{iA}$, which may contain industry classification based factors, style factors, principal component based factors, {\em etc.} -- {\em a priori} it can be arbitrary. A cruder version of the Russian-doll approximation then is that the factor covariance matrix $\Phi_{AB}$ is diagonal: $\Phi_{AB} \approx\zeta_A^2~\delta_{AB}$. However, as before, even with this approximation there is no guarantee that all $\xi_i^2 > 0$, where $\xi_i$ is specific risk. As in Section \ref{sec4}, instead of modeling the covariance matrix $C_{ij}$ via a factor model, we can {\em ad hoc} approximate the correlation matrix $\Psi_{ij}$ via a factor model ($C_{ij} = \sqrt{C_{ii}}\sqrt{C_{jj}}~\Psi_{ij}$):
\begin{eqnarray}
 \Psi_{ij}\approx \Gamma_{ij} \equiv \xi_i^2~\delta_{ij} + \sum_{A=1}^K \zeta_A^2~\Omega_{iA}~\Omega_{jA}
\end{eqnarray}
In the binary case $\Omega_{iA} = \delta_{G(i),A}$ we have the conditions
\begin{equation}
 \xi_i^2 + \zeta_{G(i)}^2 = 1
\end{equation}
so, as above, we can apply a simple Ansatz, {\em e.g.}, $\xi_i^2 = \zeta_A^2 = 1/2$. However, for non-binary $\Omega_{iA}$ a simple Ansatz does not exist and the aforementioned proprietary algorithms are required. This is also the case when we have a combination of (a small number of) style factors and binary industry factors.

\subsection*{Acknowledgments}

{}The bulk of this work was completed during my visit at the Institute for Advanced Study at Hong Kong University of Science and Technology. I would like to thank the IAS and its director Prof. Henry Tye for their hospitality. I would also like to thank Jean-Philippe Bouchaud and R{\'e}my Chicheportiche for correspondence on their use of nested factor models for modeling non-linear dependencies in stock returns.

\begin{table}[ht]
\caption{Simulation results for the 4 weighted regression alphas plus the optimized alpha discussed in Section \ref{sec4}.} 
\begin{tabular}{l l l l} 
\\
\hline\hline 
Alpha & ROC & SR & CPS\\[0.5ex] 
\hline 
Regression: Intercept only & 33.59\% & 5.59 & 1.38\\
Regression: BICS Sectors & 39.28\% & 7.05 & 1.61\\
Regression: BICS Industries & 42.66\% & 8.19 & 1.75\\
Regression: BICS Sub-industries & 45.25\% & 9.22 & 1.84\\
Optimization: $\xi_i^2=\zeta_A^2 =\eta_a^2 = \sigma_\alpha^2 = X = 1/5$ & 46.31\% & 9.49 & 1.90\\[1ex] 
\hline 
\end{tabular}
\label{table1} 
\end{table}

\begin{table}[ht]
\noindent
\caption{Simulation results for Table 1 regressions with unit regression weights.\,\,\,\,\,\,\,\,\,\,\,\,} 
\begin{tabular}{l l l l} 
\\
\hline\hline 
Alpha & ROC & SR & CPS\\[0.5ex] 
\hline 
Regression: Intercept only & 38.64\% & 5.14 & 1.01\\
Regression: BICS Sectors & 44.58\% & 6.21 & 1.17\\
Regression: BICS Industries & 49.00\% & 7.15 & 1.29\\
Regression: BICS Sub-industries& 51.77\% & 7.87 & 1.36\\ [1ex] 
\hline 
\end{tabular}
\label{table2} 
\end{table}

\newpage
\begin{figure}[ht]
\centerline{\epsfxsize 4.truein \epsfysize 4.truein\epsfbox{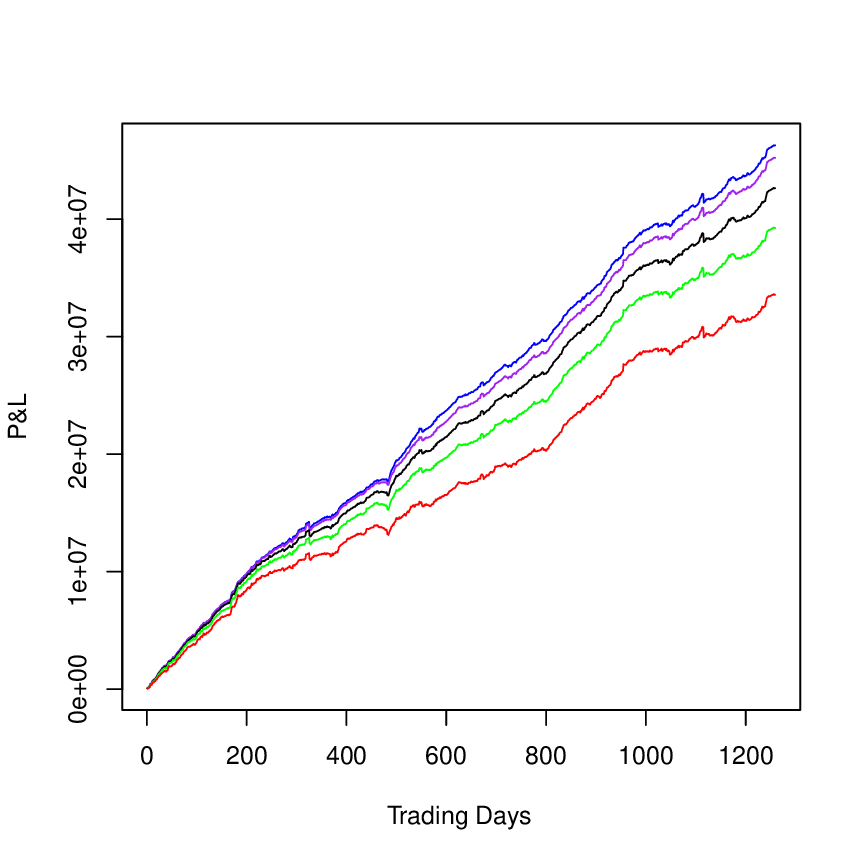}}
\noindent{\small {Figure 1. P\&L graphs for the intraday alphas discussed in Section \ref{sec4}, with a summary in Table \ref{table1}. Bottom-to-top-performing: i) Weighted regression over intercept only, ii) weighted regression over BICS sectors, iii) weighted regression over BICS industries, iv) weighted regression over BICS sub-industries, and v) optimization using the approximate Russian-doll model (\ref{crude}). The investment level is \$10M long plus \$10M short.}}

\end{figure}

\end{document}